\documentclass[aps,pra,twocolumn,superscriptaddress]{revtex4-1}
\usepackage{graphicx}
\usepackage{amsmath}
\usepackage{braket}
\usepackage[utf8]{inputenc}
\usepackage[T1]{fontenc}
\usepackage{hyperref}
\usepackage{siunitx}
\DeclareSIUnit\gauss{G}

\begin{document}

\title{Anomalous loss behavior in a single-component Fermi gas close to a $p$-wave Feshbach resonance}
\author{K. Welz}
\email{These authors contributed equally to this work.}
\affiliation{Physikalisches Institut, Universität Heidelberg, Im Neuenheimer Feld 226, 69120, Heidelberg, Germany}

\author{M. Gerken}
\email{These authors contributed equally to this work.}
\affiliation{Physikalisches Institut, Universität Heidelberg, Im Neuenheimer Feld 226, 69120, Heidelberg, Germany}

\author{B. Zhu}
\affiliation{HSBC Lab-China, Guangzhou 510620, China}

\author{E. Lippi}
\affiliation{Physikalisches Institut, Universität Heidelberg, Im Neuenheimer Feld 226, 69120, Heidelberg, Germany}

\author{M. Rautenberg}
\affiliation{Physikalisches Institut, Universität Heidelberg, Im Neuenheimer Feld 226, 69120, Heidelberg, Germany}

\author{L. Chomaz}
\email{chomaz@physi.uni-heidelberg.de}
\affiliation{Physikalisches Institut, Universität Heidelberg, Im Neuenheimer Feld 226, 69120, Heidelberg, Germany}

\author{M. Weidemüller}
\email{weidemueller@uni-heidelberg.de}
\affiliation{Physikalisches Institut, Universität Heidelberg, Im Neuenheimer Feld 226, 69120, Heidelberg, Germany}

\date{\today}

\begin{abstract}

We theoretically investigate three-body losses in a single-component Fermi gas near a $p$-wave Feshbach resonance in the interacting, non-unitary regime.
We extend the cascade model introduced by Waseem \textit{et al.} [M. Waseem, J. Yoshida, T. Saito, and T. Mukaiyama, Phys. Rev. A \textbf{99}, 052704 (2019)] to describe the elastic and inelastic collision processes. 
We find that the loss behavior exhibits a $n^3$ and an anomalous $n^2$ density dependence for a ratio of elastic-to-inelastic collision rate larger and smaller than 1, respectively.
The corresponding evolutions of the energy distribution show collisional cooling or evolution toward low-energetic non-thermalized steady states, respectively.
These findings are particularly relevant for understanding atom loss and energetic evolution of ultracold gases of fermionic lithium atoms in their ground state.

\end{abstract}


\maketitle

\section{\label{Introduction} Introduction}

Magnetic Feshbach resonances are essential for tuning the collisional properties of ultracold atomic and molecular gases~\cite{RevModPhys.82.1225, RevModPhys.80.885}.
The minimal model of Feshbach resonances takes into account two coupled collision channels, the scattering  channel and a closed channel with a bound state whose energy lies, and can be tuned around, the scattering threshold of the open channel. 
Together with the resonant behavior of the elastic binary collisions are also inelastic collisions are altered near magnetic Feshbach resonances.  
Understanding inelastic collisions and the related losses near such resonances has been essential for their exploitation.
As a paradigmatic example, the stability of a degenerate two-spin Fermi sample, caused by the suppression of possible loss processes~\cite{PhysRevA.67.010703, PhysRevLett.93.090404},
led first to the production of a Bose-Einstein condensate (BEC) of Fermi dimers~\cite{PhysRevLett.92.120401}
and later to the investigation of the BEC-Bardeen-Cooper-Schrieffer (BCS) crossover~\cite{ PhysRevLett.93.050401, PhysRevLett.92.203201, PhysRevLett.95.020404}.

In the ultracold regime, collisions in the $s$-partial wave typically dominate, see, e.g.,~\cite{RevModPhys.82.1225, RevModPhys.80.885}.
However, in the case of fermions, $s$-wave collisions are precluded between identical particles.
Feshbach resonances enhancing the $p$-wave scattering (so-called $p$-wave Feshbach resonances) were observed~\cite{regal2003tpw, PhysRevA.70.030702} between fermions of the same spin state.
Interest in resonantly $p$-wave interacting fermions arises, in particular, from the possibility of a rich quantum phase diagram involving anisotropic $p$-wave superfluidity~\cite{PhysRevLett.94.230403, PhysRevLett.95.070404}.
Such phases, however, require particularly low temperatures.
Driven by this interest, great theoretical and experimental effort has been devoted to understanding the elastic and inelastic scattering processes close to $p$-wave Feshbach resonances~\cite{regal2003tpw, PhysRevA.70.030702, PhysRevA.71.062710, PhysRevA.88.012710, PhysRevLett.120.133401, PhysRevA.98.020702}.
For the case of two-component Fermi gases with resonant $s$-wave interactions, inelastic processes of a two-body nature, induced by dipole-dipole interactions, dominate~\cite{ PhysRevA.70.030702, PhysRevA.96.062704}.
A single-component Fermi gas prepared in the lowest hyperfine state suppresses spin-flip relaxation such that the dominant inelastic process should involve three atoms~\cite{regal2003tpw, PhysRevA.70.030702, PhysRevLett.90.053202, PhysRevLett.120.133401, PhysRevA.98.020702}.

One possibility for the scattering of three $p$-wave interacting fermions is the direct collision of three atoms leading to their recombination into a deeply bound dimer and one atom, with a large amount of energy released~\cite{PhysRevLett.90.053202}.
However, as mentioned above, this effect is strongly suppressed due to Fermi statistics and instead, in some regimes, it was hypothesized that the dominant process for the scattering of three atoms involves an intermediate stage, in which a weakly-bound dimer is created before colliding with a third atom. 
To describe losses in the latter regime and explain experimental observations in ground-state $^6\textrm{Li}$ atoms in the vicinity of a $p$-wave Feshbach resonance, Waseem \textit{et al.}~\cite{PhysRevA.99.052704} used a cascade model inspired from earlier works in Bose-condensed and thermal gases~\cite{TIMMERMANS, PhysRevA.60.R765,PhysRevA.67.050701}.
In this model, the inelastic process is split into two steps.
First, two atoms form a weakly bound dimer via elastic scattering.
Second, the dimer is vibrationally quenched to a deeply bound state by inelastic collision with an observer atom, resulting in the loss of the three atoms.
Alternatively, the weakly bound dimer can break up into two free atoms, leading to an overall elastic scattering event.

In this paper we present an extension of the cascade model studied in Waseem \textit{et al.}~\cite{PhysRevA.99.052704} by investigating the two extreme regimes of the ratio of the rates of elastic-to-inelastic collisions.
In the first regime where elastic collisions dominate over inelastic collisions, we find, as in Ref.~\cite{PhysRevA.99.052704}, that the atom-loss behavior remains that of a three-body process.
Furthermore, we investigate the temperature and phase-space density evolution and we identify the regimes of collisional heating and collisional cooling as a function of the ratio of the dimer binding energy to the mean thermal energy.
The possibility of collisional cooling is similar to earlier findings, see, e.g., Refs~\cite{PhysRevA.80.030702, PhysRevA.91.043626, PhysRevA.87.053608, 235386, peng2021cooling}.
In the second regime where inelastic collisions dominate, we find that the loss dynamic distinctly follows that of an effective two-body process.
Due to the slow  thermalization compared to the losses, and due to the energy dependence of the collision processes, the momentum distribution is found to evolve to a low-energy non-thermal steady state.

The paper is organized as follows.
In Sec.~\ref{Three-body}, we describe the elastic and inelastic collisions via a cascade model and explain the basic assumptions needed for the model and for separating the two loss behavior regimes.
Atom loss and temperature evolution in the thermalized regime are described in Sec. ~\ref{Thermalized}.
The non-thermalized regime with an extension to an energy-dependent model of the losses is studied in Sec.~\ref{Non-Thermalized}.
We conclude in Sec.~\ref{Discussion}.

\section{\label{Three-body} Collisional cascade model}

\begin{figure}
    \hspace{-0.4cm}
    \includegraphics[width = 0.5\textwidth]{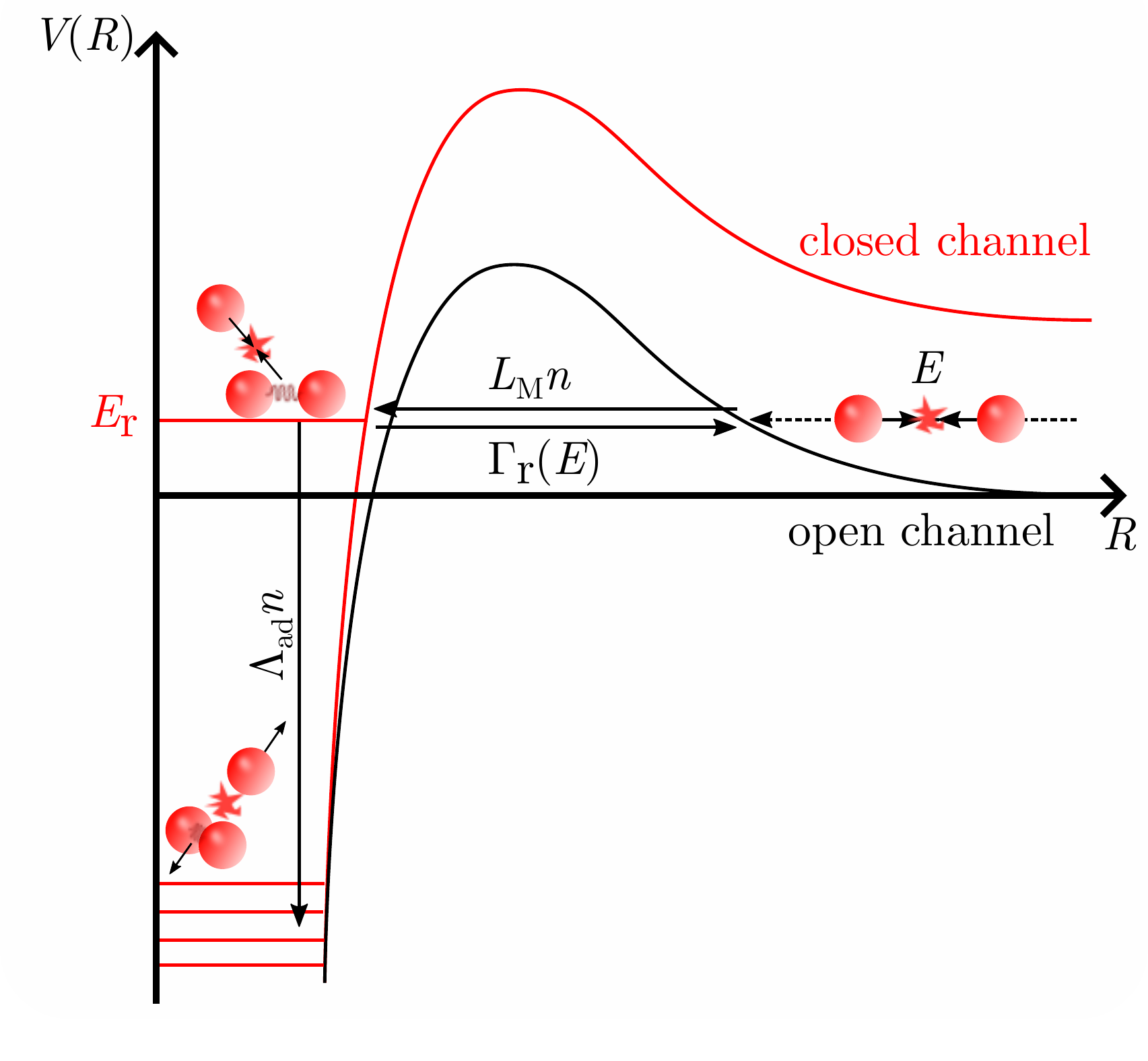}
    \caption{\label{Fig1} Schematic depiction of the $p$-wave scattering process for two-channel scattering. The figure shows the potential of the incoming scattering atoms at energy $E$ and the closed channel with a resonant dimer state at energy $E_\textrm{r}$. Incoming atoms can tunnel through the centrifugal barrier and create weakly bound dimers at a rate $L_\textrm{M} n$, with $n$ the density of the atoms. Weakly bound dimers can, in return, decay into two free atoms with rate $\Gamma_{\textrm{r}}(E)$. This elastic scattering process leads to energy redistribution, i.e., thermalization. When weakly bound dimers collide with a free atom, they collapse into a deeply bound dimer at a rate $\Lambda_\textrm{ad}n$. This inelastic process leads to a loss since the binding energy converted to kinetic energy is large compared to the trap depth of the atomic trap.}
\end{figure}

In our model, we consider free identical fermions colliding in a $p$-wave open channel and a unique weakly bound dimer state of energy $E_\textrm{r}$ in a closed channel that induces a resonant behavior of the scattering, see Fig.~\ref{Fig1}.
The elastic and inelastic collisions are each described by a cascade of two processes, as in Ref.~\cite{PhysRevA.99.052704}.
Both cascades start with the formation of a weakly bound dimer in the collision of two free atoms.
In the elastic collision case, the second process is the break-up of this dimer back into two free atoms, see the solid arrow pointing to the right in Fig.~\ref{Fig1}.
Here, a redistribution of kinetic energy between the colliding atoms can occur.

In the inelastic collision case, instead, the second process is the collision of the weakly bound dimer with a free atom and its relaxation into a deeply bound dimer, see the vertical arrow in Fig.~\ref{Fig1}.
In the later process, a third atom is necessary because of energy and momentum conservation.
With the relaxation to the deeply bound state, both the atom and dimer acquire large kinetic energy such that they leave the trap and are irreversibly removed from the system of interest.
As a consequence, the relaxation is an irreversible process in contrast to the creation and break-up of the weakly bound dimer.

The inelastic and elastic cascades compete and their dynamics can be described with two coupled differential equations describing the evolution of the density of free atoms, $n$, and that of weakly bound dimers, $n_{\textrm{D}}$~\cite{PhysRevA.99.052704, PhysRevLett.120.193402} as follows:
\begin{align}
    \label{eq3}
    \frac{d n_{\textrm{D}}}{dt} &= - (\Gamma_\textrm{r} + \Lambda_{\textrm{ad}} n) n_\textrm{D} + L_{\textrm{M}} n^2,\\
    \label{eq4}
    \frac{d n}{dt} &=  -( 2 L_{\textrm{M}} n + \Lambda_{\textrm{ad}} n_\textrm{D}) n + 2 \Gamma_\textrm{r} n_\textrm{D}.
\end{align}
For both equations, the term with rate coefficient $L_\textrm{M}$ describes the creation of a weakly bound dimer from two free atoms.
The term with rate $\Gamma_\textrm{r}$ corresponds to the breakup of a weakly bound dimer.
The atom-dimer collisions are described by a rate coefficient $\Lambda_\textrm{ad}$, yielding losses from the system.
The bracketed contributions in Eqs.~\eqref{eq3} and \eqref{eq4} denote the atomic and dimer density decay rates, respectively.
These equations neglect other loss processes such as the collision with background gases, direct three-body recombination, dimer-dimer relaxations, and secondary collisions, which are assumed to be slow compared to the investigated processes~\cite{PhysRevLett.120.193402, PhysRevA.60.R765}.
The validity of these approximations is discussed in the specific settings of the $159~\si{\gauss}$ $p$-wave Feshbach resonance of $^6\textrm{Li}$ and with the parameters of Ref.~\cite{PhysRevA.99.052704} in Appendix~\hyperref[appendixA]{A}.

We now aim to relate the rate coefficients relevant for the cascade model to the parameters describing the scattering close to a $p$-wave resonance.
For two atoms of mass $m$ and velocities $\boldsymbol{v}$ and $\boldsymbol{v'}$, the relative collisional energy is given by $E = m |\boldsymbol{v}-\boldsymbol{v'}|^2/4 = m {v}_{\rm rel}^2/4 = \hbar^2 k^2/m$.
The $p$-wave scattering phase shift $\delta_\textrm{p}(k)$ can be described in the picture of two-channel scattering by an effective range expansion: $k^3 \cot\delta_\textrm{p}(k) = -V_\textrm{p}^{-1} - k_\textrm{e} k^2$, where $k_\textrm{e}$ is the effective range of the potential~\cite{PhysRevA.95.032710} and $V_\textrm{p}= V_\textrm{bg} \left[1-\Delta B/(B-B_\textrm{0})\right] \approx -V_\textrm{bg} \Delta B/(B-B_\textrm{0})$ is the magnetic-field-dependent scattering volume.
$V_\textrm{bg}$ is the background-scattering volume, $\Delta B$ is the resonance width, and $B-B_\textrm{0}$ the magnetic field detuning from the Feshbach resonance at $B_\textrm{0}$.
The $p$-wave scattering amplitude is then given by $f_\textrm{p}(k) = k^2/\left[ k^3 \cot\boldsymbol{(}\delta_\textrm{p}(k)\boldsymbol{)} - ik^3\right]$, in the limit of $|V_\textrm{p}|^{-1} \ll k_\textrm{e}^3$, and diverges for the energy:
\begin{equation}
    \label{eq1}
    E_{\textrm{pole}} = \frac{-\hbar^2}{m V_\textrm{p} k_\textrm{e}} - i \frac{ 2 \sqrt{m} E^{3/2}}{k_\textrm{e} \hbar}.
\end{equation}
From the real part of Eq.~\eqref{eq1} we deduce that the molecular bound state exists for $V_\textrm{p}<0$ %at the resonance position 
and has an energy 
\begin{equation}
    \label{Er}
    E_\textrm{r} = \frac{\hbar^2}{m |V_\textrm{p}| k_\textrm{e}} \equiv \frac{\hbar^2 k_\textrm{r}^2}{m} \approx \frac{\hbar^2(B-B_0)}{m |V_\textrm{bg}\Delta B| k_\textrm{e}}.
\end{equation}
From the imaginary part of Eq.~\eqref{eq1} we infer that its energy-dependent break-up rate in two free atoms of relative energy $E$, is given by $\Gamma_\textrm{r}(E) = 2 \sqrt{m} E^{3/2}/(k_\textrm{e} \hbar^2)$~\cite{GURARIE20072}.
It is interesting to note that $\hbar \Gamma_\textrm{r}(E_\textrm{r})=2 E_\textrm{r}/\sqrt{|V_p| k_e^3} \ll E_\textrm{r}$ holds
in the limit $|V_\textrm{p}|^{-1} \ll k_\textrm{e}^3$ considered above.
The energy-dependent rate coefficient at which two atoms of relative collisional energy $E$ form a weakly-bound dimer is given by~\cite{PhysRevA.80.030702}
\begin{equation}
    \label{eq2}
    K_{\textrm{M}}(E) = v_\textrm{rel} \frac{\pi}{k^2} \frac{3\ \hbar^2 \Gamma_\textrm{r}(E)^2}{(E-E_\textrm{r})^2 + \hbar^2 \Gamma_\textrm{r}(E)^2/4}
\end{equation}
and depends on the resonance width $\Gamma_\textrm{r}(E)$ and the resonance energy $E_\textrm{r}$.

In this work, we focus on the regime where the cascade model explains the dominant loss process, in contrast to the non-interacting regime and the unitary regime where direct three-body recombination models were applied~\cite{PhysRevLett.90.053202}.
This regime is defined by $E_\mathrm{r} \gtrsim 3k_\textrm{B} T/2$ and distinguishes itself from the  unitary (non-interacting) regime via $3k_\mathrm{B} T/2 > E_\textrm{r}$ ($k_\textrm{B} T < E_\mathrm{r}/10$)~\cite{PhysRevA.99.052704, PhysRevLett.120.133401} where $T$ is the gas temperature and $k_\textrm{B}$ is the Boltzmann constant.
In the interacting non-unitary regime of interest, we can thus assume $\hbar \Gamma_\textrm{r}(E_\textrm{r})\ll k_B T$ using the limit of $|V_\textrm{p}|^{-1} \ll k_\textrm{e}^3$.
In this case, we can approximate the dimer formation rate coefficient by a Dirac delta distribution, 
\begin{equation}
    \label{KME}
    K_{\textrm{M}}(E) \approx \Gamma_\textrm{r}(E_\textrm{r}) \frac{3}{2 \pi} \left( \frac{2 \pi}{k_\textrm{r}} \right)^3 \delta \left( 1 - E/E_\textrm{r} \right).
\end{equation}
This implies that two colliding atoms must have a relative energy matching $E_\textrm{r}$ to be able to form a dimer.
As a consequence, we can approximate the collision rates as energy independent as done in Eqs.~\eqref{eq3} and \eqref{eq4} with the dimer break-up rate 
\begin{equation}
    \label{Gammar}
    \Gamma_\textrm{r} \equiv \Gamma_\textrm{r}(E_{\textrm{r}}) = \frac{2 \sqrt{m} E_{\textrm{r}}^{3/2}}{k_\textrm{e} \hbar^2}
\end{equation}
and the dimer creation rate coefficient defined by a thermal averaging of $K_{\textrm{M}}(E)$, 
\begin{equation}
    \label{LM}
    L_{\textrm{M}} = 3 \Gamma_\textrm{r} \left(\frac{6 \pi }{k_\textrm{T}^2}\right)^{3/2} e^{-E_\textrm{r}/(k_\textrm{B} T)},
\end{equation} with $k_\textrm{T} = \sqrt{3 m k_\textrm{B} T/2}/\hbar$ the thermal wave number. Both rates are magnetic-field dependent via Eq.~\eqref{Er}.
The vibrational quenching rate coefficient $\Lambda_\textrm{ad}$, quantifying the atom-dimer collision rate, is taken to be constant and independent of the dimer's and atom's momenta, see also Appendix~\hyperref[appendixA]{A}.

Under the assumption of a thermal gas in the interacting non-unitary regime we deduce that the dimer creation rate coefficient $L_{\textrm{M}}$ in Eq.~\eqref{LM} is much smaller than the dimer break-up rate $\Gamma_\textrm{r}$ in Eq.~\eqref{Gammar}.
Furthermore, it is reasonable to assume that the initial dimer density is small compared to the initial atom density.
This results in the rate $(\Gamma_\textrm{r} + \Lambda_\textrm{ad} n)$ in Eq.~\eqref{eq3} at which the dimer density decreases to be much larger than the decay rate of the atom density $(2 L_\textrm{M} n + \Lambda_\textrm{ad} n_\textrm{D})$ in Eq.~\eqref{eq4}.
Therefore, we can conclude that the dimer density is always in a steady state compared to the atomic density, such that $\frac{d n_\textrm{D}}{d t}=0$ on the scale of the atom loss and
\begin{equation}
    \label{eqnD}
    n_\textrm{D} = \frac{L_\textrm{M} n^2}{n \Lambda_{\textrm{ad}} + \Gamma_\textrm{r}}.
\end{equation}
The density-loss equation can then be rewritten as:
\begin{equation}
    \label{eq5}
    \frac{d n}{dt} = - \frac{9 \Lambda_{\textrm{ad}} \Gamma_\textrm{r}}{n \Lambda_{\textrm{ad}} + \Gamma_\textrm{r}} \left( \frac{6 \pi}{k_\textrm{T}^2} \right)^{3/2} e^{-E_\textrm{r}/(k_\textrm{B} T)}n^3.
\end{equation}

In Appendix~\hyperref[appendixA]{A}, we extract the scattering and loss parameters for a $p$-wave Feshbach resonance (close to $B=159~\si{\gauss}$) of $^6\textrm{Li}$ in the spin state $\ket{F, m_\textrm{F}} = \ket{1/2,1/2}$ using the measurements of Ref.~\cite{PhysRevA.99.052704}.
We demonstrate that the interacting non-unitary regime can be reached experimentally while still being able to change the ratio $\Gamma_\textrm{r}/(n_\textrm{0}\Lambda_{\textrm{ad}})$.
We also self-consistently check the validity of the approximations made in our model in this specific system.
In particular, the average dimer density is evaluated to be at least one order of magnitude smaller than the atom density. Based on this estimate, we confirm that other loss processes such as dimer-dimer relaxation and secondary collisions can be neglected.

In the following, we investigate the consequences of Eq.~\eqref{eq5} in two limiting scenarios leading to the separation into a thermalized and a non-thermalized regime.

\section{\label{Results} Results}

\subsection{\label{Thermalized} Thermalized regime}

\begin{figure}
    \includegraphics[width = 0.48\textwidth]{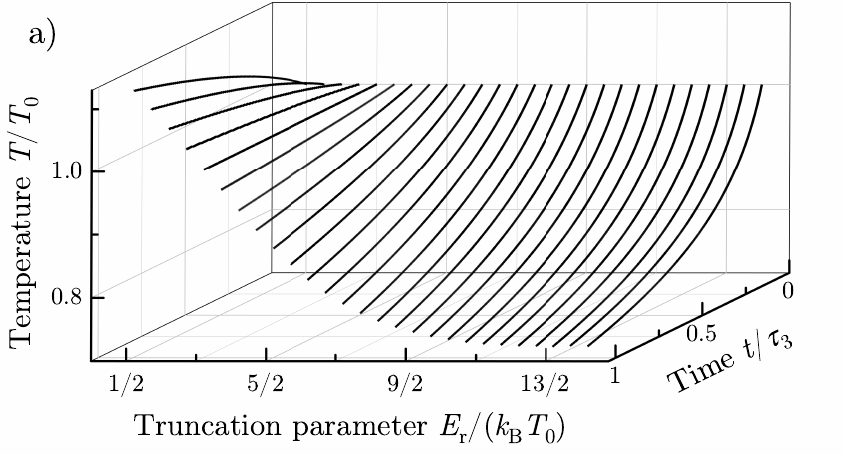}
    \includegraphics[width = 0.48\textwidth]{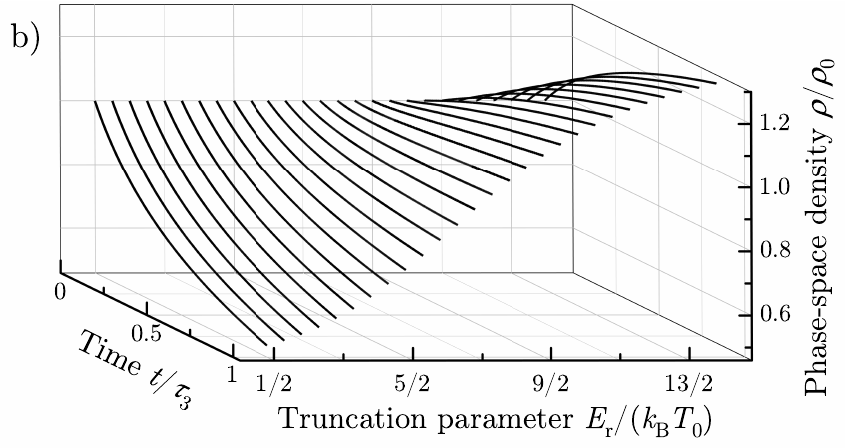}
    \caption{\label{Fig2} (a) The temperature $T$ in units of the initial temperature $T_\textrm{0} = T(t=0)$ and (b) the phase-space density $\rho$ in units of $\rho_\textrm{0} = \rho(t=0)$ as a function of time t and the truncation parameter $E_\textrm{r}/(k_\textrm{B} T_\textrm{0})$.
    The temperature decreases for $\eta_\textrm{0}>3/2$ and the phase-space density increases for $\eta_\textrm{0}>9/2$.}
\end{figure}

We start by investigating the regime where
\begin{equation}
    \label{inequelityA}
    \Gamma_{\textrm{r}}\gg n\Lambda_{\textrm{ad}}.
\end{equation}
In this case the creation and breakup of weakly bound dimers happens quickly compared to the relaxation into a deeply bound dimer.
Considering the two scattering outcomes, this condition implies a large elastic to inelastic collision ratio, and therefore a thermalized system.

In this thermalized regime, Eq.~\eqref{eq5} simplifies to 
\begin{equation}
    \label{eqn3}
    \frac{dn}{dt}=-L_\textrm{3}n^3,
\end{equation}
with the inelastic scattering process yielding an effective three-body decay of the density with the three-body loss coefficient $L_\textrm{3} = 9 \Lambda_{\textrm{ad}} (6 \pi/k_\textrm{T}^2 )^{3/2} e^{-E_\textrm{r}/(k_\textrm{B} T)}$.
$L_\textrm{3}$ only depends on $\Lambda_{\textrm{ad}}$ and not on $\Gamma_\textrm{r}$ evidencing that the atom-dimer collision process is the limiting effect in the loss.

In the temperature evolution during the loss process in the thermalized regime, either heating or cooling can occur.
For simplicity we assume a cloud trapped in a uniform box potential of volume $V$.
We further assume that the thermalization is instantaneous compared to the loss events, as justified by the inequality~\eqref{inequelityA}.
We first consider a single cascaded loss event where the atoms $1$ and $2$ form a dimer before colliding with atom $3$.
We denote as $\boldsymbol{v}_i$ the velocity of atom $i$.
Given the assumption of the interacting non-unitary regime, $\boldsymbol{v}_\textrm{1}$ and $\boldsymbol{v}_\textrm{2}$ satisfy $\frac{m}{4} |\boldsymbol{v}_\textrm{1} - \boldsymbol{v}_\textrm{2}|^2 =E_\textrm{r}$.
Assuming a temperature $T$ of the gas before the collision, the ensemble-averaged (denoted $\langle \cdot \rangle$) values satisfy $\frac{m}{4} \langle |\boldsymbol{v}_\textrm{1} + \boldsymbol{v}_\textrm{2}|^2 \rangle = \frac{m}{2} \langle |\boldsymbol{v}_\textrm{3}|^2 \rangle =3k_\textrm{B} T/2$.
Therefore the average total energy lost when the three atoms leave the cloud is $\langle E_\textrm{loss} \rangle = \sum_{i=1}^3 \frac{m}{2} \langle |\boldsymbol{v}_i|^2 \rangle= E_\textrm{r} + 3 k_\textrm{B} T$.
In turn, the change in temperature in the gas induced by a single loss event $\delta T_\textrm{loss}$ is deduced from the average energy lost and the average kinetic energy of the three atoms via $\frac{3}{2}Nk_\textrm{B} \delta T_\textrm{loss} = \frac{9}{2} k_\textrm{B} T - \langle E_\textrm{loss} \rangle = \frac{3}{2} k_\textrm{B} T - E_\textrm{r}$ with $N$ the  number of atoms in the gas after the loss.
As in a box trap $n=N/V$, the rate of total loss events is given by $L_\textrm{3}n^2N/3$.
This results in
\begin{equation}
    \label{eq7}
    \frac{d T}{d t} = L_\textrm{3} n^2 \frac{2T}{9} \left( \frac{3}{2} - \eta \right),
\end{equation}
with the truncation parameter $\eta = E_\textrm{r}/(k_\textrm{B} T)$.
Depending on the value of $\eta$, the temperature variations are found to change sign.
For a small dimer energy such that $\eta<3/2$, low energetic particles are preferentially lost, resulting in an overall heating of the gas.
In contrast, for a large dimer energy such that $\eta>3/2$, high energy particles are preferentially lost and a cooling of the gas prevails.
For a dimer energy such that $\eta = 3/2$, the temperature of the gas is constant because the average kinetic energy of an atom lost is $3/2k_\textrm{B} T$.
Note that the regime $\eta<3/2$ actually matches the unitary regime (see Sec.~\ref{Three-body}) and in the interacting non-unitary regime of interest $\eta>3/2$ and loss-induced cooling prevails.

For the evolution with time $t$ of a gas, the initial parameters are defined as $\eta_\textrm{0} = E_\textrm{r}/(k_\textrm{B} T_\textrm{0})$, $T_\textrm{0} = T(t=0)$, and $n_\textrm{0}=n(t=0)$.
To formulate the time as unitless, we use the characteristic loss time $\tau_\textrm{3} = 1/\textbf{(}L_\textrm{3}(t=0)n_\textrm{0}^2\textbf{)}$ (typical time for $N/3$ loss events).
Figure~\ref{Fig2}~(a) shows the numerical results of the time evolution of the temperature following Eq.~\eqref{eq7} as a function of $\eta_\textrm{0}$.
As anticipated above, for $\eta_\textrm{0} = 3/2$, the temperature of the gas is constant.
For  $\eta_\textrm{0} < 3/2$, the temperature increases, with an overall change over $\tau_\textrm{3}$ of $T(\tau_\textrm{3})/T_0= 1.12$ for $\eta_\textrm{0}= 1/2$, and for  $\eta_\textrm{0} > 3/2$, the temperature decreases, with $T(\tau_\textrm{3})/T_0$ reaching $0.74$ for $\eta_\textrm{0}= 9/2$.
The speed of the temperature change saturates for $\eta_\textrm{0}>11/2$, limiting a possible cooling scheme on the characteristic loss timescale.
This is because the fast temperature decrease slows down further losses.

Considering the change in density and temperature, the change in phase-space density $\rho=n (2 \pi \hbar)^3/\sqrt{2 \pi m k_\textrm{B} T}^3$ is given by~\cite{PhysRevA.91.043626, peng2021cooling}: 
\begin{equation}
    \label{eq7.1}
    \frac{d \rho}{d t} = L_\textrm{3} n^2 \frac{\rho}{3} \left( \eta - \frac{9}{2} \right).
\end{equation}
As seen in Fig.~\ref{Fig2} (b), the time derivative of the phase-space density is found to change sign at $\eta=9/2$.
For $\eta<9/2$ ($\eta>9/2$), $\rho$ decreases (increases) with time.
The energy-dependent cascaded loss close to $p$-wave Feshbach resonances can therefore be an eligible method of collisional cooling.
The evaporation efficiency $\gamma=-\ln{(\rho/\rho_0)} /\ln{(N/N_0)}$~\cite{KETTERLE1996181} in this case is $\gamma=\eta/3-3/2$.

We note that the simple scaling laws extracted above hold for a uniform gas.
In the presence of a harmonic trap, the density is inhomogeneous and a spatial dependence of the loss behavior is expected.
For the energy-independent three-body recombination, this is known to yield a so-called “anti-evaporation” effect, i.e., loss-induced heating~\cite{PhysRevLett.91.123201}.
We foresee that a spatially dependent loss behavior may temper the cascade-induced collisional cooling discussed above.

%-------------------------------------------------

\subsection{\label{Non-Thermalized} Non-thermalized regime}

\begin{figure}
    \includegraphics[width = 0.5\textwidth]{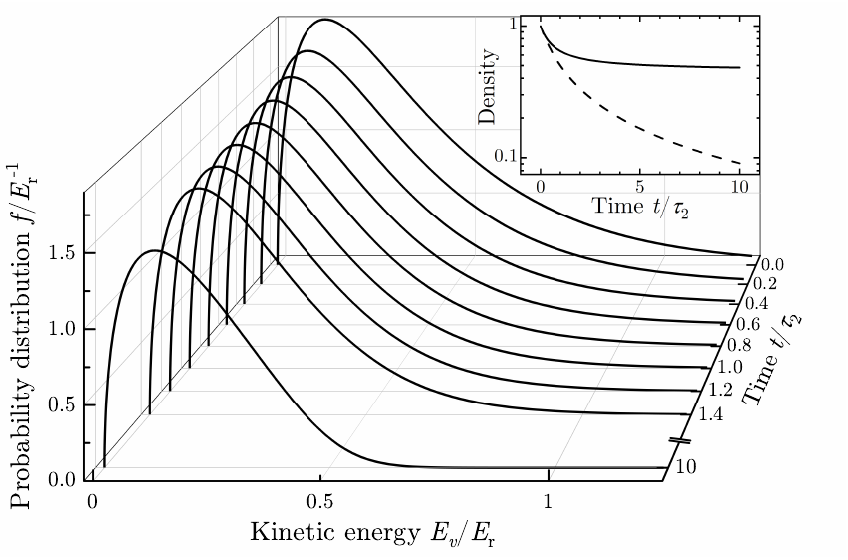}
    \caption{\label{Fig3} The kinetic energy distribution $f (E_v)$ of a gas in the non-thermalized regime in units of $E_\textrm{r}^{-1}$ for a truncation parameter $\eta_\textrm{0}=4$ after a hold time $t$ in units of the loss timescale $\tau_\textrm{2}$ in a box potential.
    Initially, $f$ is a normalized Maxwell-Boltzmann distribution of temperature $E_\textrm{r}/(4 k_\textrm{B})$ and the three-body losses lead to a decrease and reshaping of the kinetic energy distribution.
    After the axis break, at $t = 10\tau_\textrm{2}$, the distribution does not have an exponential tail of energies $E_v>E_\textrm{r}/2$ anymore.
    The inset shows the density in units of the initial density changing due to a decrease of the kinetic energy distribution (solid line) and the projected $n^2$-dependent loss for a gas of constant kinetic energy distribution (dashed line).}
\end{figure}

In the following, we consider the opposite extreme case of $\Gamma_\textrm{r} \ll n \Lambda_{\textrm{ad}}$, i.e. the breakup of the weakly bound dimers being negligible compared to the loss process.
The relaxation into the deeply bound dimer is instantaneous subsequent to the dimer creation.
This leads to a large inelastic to elastic collision ratio, and therefore a non-thermalized, out-of-equilibrium sample.
In this regime, Eq.~(\ref{eq5}) simplifies to
\begin{equation}
    \label{eq:Lossn2}
    \frac{dn}{dt}=-L_2 n^2,
\end{equation}
where the inelastic scattering process only depends on $\Gamma_{\textrm{r}}$, with $L_2 = 9 \Gamma_\textrm{r} \left(6 \pi/k_\textrm{T}^2 \right)^{3/2} e^{-E_\textrm{r}/(k_\textrm{B} T)}$.
This shows a qualitative change in loss behavior from an $n^3$ dependence in the thermalized regime to an $n^2$ dependence in the non-thermalized regime.
This change is due to the losses being limited by the number of dimers instead of the number of dimer-atom pairs.

Due to the absence of thermalization, the reshaping of the kinetic energy distribution from an initially thermalized sample is determined by the energy-dependent losses.
The change in the kinetic energy distribution is described by a differential equation obtained via the extension of Eq.~(\ref{eq4}) into a velocity dependent description (see Appendix~\hyperref[appendixB]{B}).
Figure~\ref{Fig3} shows the numerical results of the change of the kinetic energy distribution for $\eta_0=4$.
The time is expressed in units of the characteristic timescale of the losses defined by Eq.~(\ref{eq:Lossn2}), $\tau_\textrm{2} = 1/\textbf{(}L_\textrm{2}(t=0) n_\textrm{0}\textbf{)}$ (analogous to $\tau_\textrm{3}$).
At short times, the change in the energy distribution is dominated by a decrease in amplitude which evidences the atom loss.
At a later time, we observe a reshaping of the exponential tail at high energy.
In particular, the kinetic energies $E_v > E_\textrm{r}/2$ are depleted.
Finally, the loss stops after this energy tail is fully depleted, as then, no collision process bears enough relative energy to form a weakly bound dimer.
The result is a stable out-of-equilibrium kinetic energy distribution, with a depleted high-energy tail, no thermalization, and no loss.
Note that at this point, other processes that were neglected such as direct three-body recombination may still contribute.

\section{\label{Discussion} Conclusion}

% Anomalous Loss
In this work, we extended the previously studied cascade model for a single-component Fermi gas near a $p$-wave Feshbach resonance in the interacting, non-unitary regime to combine the description of thermalization and three-body loss.
We extract simple scaling laws for the time evolution of the atom number and the temperature in the thermalized case and predict anomalous loss behavior in the non-thermalized scenario.
In the later case, the atom number loss exhibits a $n^2$ dependence 
% Deviation from Gaussian momentum distribution
and a yet unexplored behavior of the sample's energy distribution: it is reshaped due to the loss with a particular depletion in the high-energy tail ($E_v \gtrsim E_\textrm{r}/2$).
Within the approximation of our model, the loss vanishes once the tail is fully depleted, leaving a low-energetic, non-thermal steady state.
% Applicability of collisional cooling
In the opposite regime, i.e., the thermalized case, we find that, in the interacting non-unitary regime [$E_\textrm{r}/(k_\textrm{B}T)>3/2$], the sample is collisionally cooled at a rate proportional to $n^2$.
This allows for an increase in phase-space density that can be controlled via the bound-state energy $E_\textrm{r}$, that is to say, in experiments via the magnetic-field value.

% Relation to other collisional cooling scenarios
It is interesting to note that cooling or energy removal via three-body loss generally occurs on different timescales as compared to standard evaporative cooling.
Furthermore, these timescales can be reduced below typical two-body evaporation timescales at low temperatures by tuning $E_\textrm{r}$.
% Comparison to evaporative cooling
In the thermalized regime, this could allow for efficient cooling of single-component Fermi gases similar to standard evaporation of $s$-wave interacting fermions.
The cooling efficiency is found to be $\gamma=\eta/3-3/2$, which can be tuned to comparable values as that of standard evaporative cooling $\gamma^\prime \gtrsim 2 \eta^\prime/3 - 1$~\cite{KETTERLE1996181} without relying on $s$-wave collisions or changing the external trapping potential.
Here $\eta^\prime$ is the standard truncation parameter, i.e., the ratio of trap depth to mean thermal energy, and is typically less than 10~\cite{KETTERLE1996181}.
%accessibility in Li experiment/ application to lithium: put later?
Finally, we find that the interacting, non-unitary regime in terms of temperature and magnetic-field detuning from the Fesh\-bach resonance is experimentally accessible using $^6\textrm{Li}$, see Appendix~\hyperref[appendixA]{A}.
Thus, our findings can be tested experimentally in future three-body loss investigations of non-unitary $p$-wave interactions.

\begin{acknowledgments}
We wish to thank Selim Jochim, Hans-Werner Hammer, Binh Tran, Tobias Krom, and Robert Freund for fruitful discussions and helpful comments.
We are also grateful to Takashi Mukaiyama and Muhammad Waseem for providing the measurement data depicted in Fig~\ref{FigA}. This work is supported by the Deutsche Forschungsgemeinschaft (DFG, German Research Foundation) Project No. 273811115 - SFB 1225 ISOQUANT and by DFG under Germany’s Excellence Strategy Frant No. EXC-2181/1 - 390900948 (Heidelberg STRUCTURES Excellence Cluster).
\end{acknowledgments}

\appendix
\section*{\label{appendixA} Appendix A: Benchmarking the cascade model}

To verify our model we compare it to the experimental data from Waseem \textit{et al}.~\cite{PhysRevA.99.052704}.
In their experiment they investigated three-body losses of $^6\textrm{Li}$ in the $\ket{F,m_\textrm{F}} = \ket{1/2,1/2}$ state close to the Feshbach resonance at $159~\si{\gauss}$ for different magnetically tuned positive binding energies $E_\textrm{r}$ on the order of the kinetic energy of the atoms $3/2 k_\textrm{B} T$.
They extracted the three-body loss parameter $L_3$ from a fit of the atom-loss curves [see Eq.~\eqref{eqn3}]. %\label{eqn3}
Then they used the cascade model to fit the rate coefficient of the vibrational quenching $\Lambda_{\textrm{ad}}$ from $L_\textrm{3}$ \footnote[1]{In \cite{PhysRevA.99.052704} the notation is $K_{\textrm{ad}}$ for the vibrational quenching rate coefficient.}.

\begin{figure}
    \includegraphics[width = 0.48\textwidth]{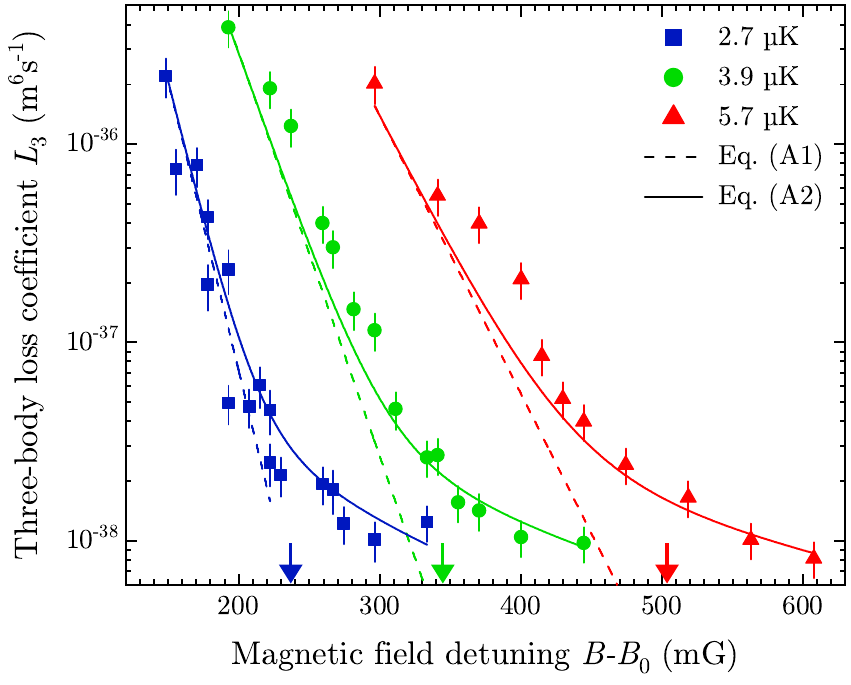}
    \caption{\label{FigA} The three-body loss coefficient $L_\textrm{3}$ close to the $p$-wave Feshbach resonance at $B_\textrm{0} = 159.17(5)~\si{\gauss}$ from the publication~\cite{PhysRevA.99.052704} for three different sets of temperatures of $2.7~\si{\micro\kelvin}$, $3.9~\si{\micro\kelvin}$ and $5.7~\si{\micro\kelvin}$ shown as squares, circles, and triangles respectively.
    The arrows mark the points where $E_\textrm{r}/(k_\textrm{B}T_\textrm{0}) =10$, separating the interacting and non-interacting regime.
    The solid curves, describing the crossover between the two regimes, are a fit of Eq.~\eqref{eqLadd} to the data of the three temperatures with a fit result for the vibrational quenching rate coefficient of $\Lambda_\textrm{ad} = 3.6(6) \times 10^{-14}~\si{\meter\cubed\per\second}$.
    The dashed curves show the loss coefficient of only the cascade model with the fit result $\Lambda_\textrm{ad}$ included in Eq.~\eqref{eqLint}.}
\end{figure}

Here we use the $L_\textrm{3}$ values fitted to their atom-loss measurements and extract $\Lambda_{\textrm{ad}}$ using the model described in this paper.
Figure~\ref{FigA} shows $L_\textrm{3}$ for three sets of measurements of samples at temperatures of $2.7~\si{\micro\kelvin}$, $3.9~\si{\micro\kelvin}$, and $5.7~\si{\micro\kelvin}$ for magnetic field detunings between $150~\si{\milli\gauss}$ and $600~\si{\milli\gauss}$.
The mean densities for the data sets are $n_\textrm{1} = 1.2 \times 10^{18}~\si{\per\meter\cubed}$, $n_\textrm{2} = 1.3 \times 10^{18}~\si{\per\meter\cubed}$, and  $n_\textrm{3} = 1.5 \times 10^{18}~\si{\per\meter\cubed}$ for the respective temperatures.

For these measurements, the gas was neither fully in the thermalized nor in the non-thermalized regime with regards to the elastic-to-inelastic collision ratio, and thus the loss coefficient should be extracted from Eq.~\eqref{eq5} for small losses with
\begin{equation}
    \label{eqLint}
    \tag{A1}
    L_\textrm{3}^\textrm{int}= \frac{9 \Lambda_{\textrm{ad}} \Gamma_\textrm{r}}{n_\textrm{0} \Lambda_\textrm{ad} + \Gamma_\textrm{r}} \left( \frac{6 \pi}{k_\textrm{T}^2} \right)^{3/2} e^{-E_\textrm{r}/(k_\textrm{B} T)}.
\end{equation}
The data in Fig.~\ref{FigA} show a transition from the interacting non-unitary regime to the non-interacting regime, making a second description of the loss necessary to explain the measurements.
Because the assumption that virtually all atoms lost have the collision energy $E_\textrm{r}$ does not hold in the non-interacting regime, the molecule creation rate cannot be approximated with a delta distribution.
Instead, in this regime, the losses are described via direct three-body recombination yielding $L_\textrm{3} \propto V_\textrm{p}^{8/3}$ and no density-dependent regimes~\cite{PhysRevLett.90.053202}.
We add the two results for $L_\textrm{3}$ to describe the crossover of the two regimes:
\begin{equation}
    \label{eqLadd}
    \tag{A2}
    L_\textrm{3}= L_\textrm{3}^\textrm{int} + C\frac{\hbar}{m}k_\textrm{T}^4V_\textrm{p}^{8/3}
\end{equation}
Here $C=2 \times 10^{6}$ is dimensionless and quantifies the coupling strength between the closed channel and the deeply bound state~\cite{PhysRevLett.120.133401}.
Equation~\eqref{eqLadd} describes the three data sets from Fig.~\ref{FigA} in the crossover regime with $\Lambda_{\textrm{ad}}$ as the only free parameter.
We fix the other parameters to the values $E_\textrm{r}/[k_\textrm{B}(B-B_\textrm{0})] = 113 \pm 7~\si{\micro\kelvin\per\gauss}$ from Fuchs \textit{et al.}~\cite{PhysRevA.77.053616} and $k_\textrm{e} = 0.055(5)~\si{\per\bohr}$ from Nakasuji \textit{et al.}~\cite{PhysRevA.88.012710} and fit Eq.~\eqref{eqLadd} to the data points.

We weight each data set equally, independent of the number of points.
The resulting curves are shown in Fig.~\ref{FigA} (solid lines), and they are in good agreement with the data.
We extract a vibrational quenching rate coefficient $\Lambda_{\textrm{ad}} = 3.6(6) \times 10^{-14}~\si{\meter\cubed\per\second}$~\footnote[3]{Due to a difference in the definition of $L_\textrm{3}$ as compared to~\cite{PhysRevA.99.052704}, our $\Lambda_{\textrm{ad}}$ is not comparable to their parameter $K_{\textrm{ad}}$.}.
By fitting the data with Eq.~\eqref{eqLadd} we reduce the relative error to $\Delta \Lambda_{\textrm{ad}}/\Lambda_{\textrm{ad}} \approx 0.17$ compared to $\Delta K_{\textrm{ad}} /K_{\textrm{ad}} \approx 0.38$ from Waseem \textit{et al.}~\cite{PhysRevA.99.052704}.
It is interesting to note that the bare cascade model of Eq.~\eqref{eqLint}, using the above-extracted $\Lambda_{\textrm{ad}}$, explains well the data for each temperature in the region where the system is close enough to the resonance (dashed lines), identifying the interacting non-unitary regime.

\begin{figure}
    \includegraphics[width = 0.48\textwidth]{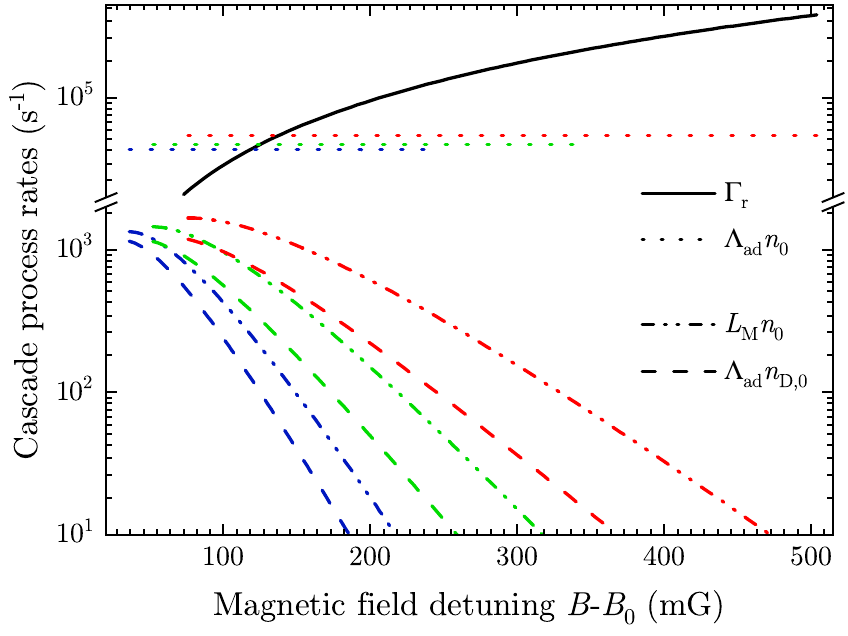}
    \caption{\label{FigB} The cascade process rates of Eqs.~\eqref{eq3} and \eqref{eq4} as a function of the magnetic field detuning $B-B_\textrm{0}$.
    The three colors correspond to the rates for the densities and temperatures of the loss measurements depicted with the same colors in Fig.~\ref{FigA}.
    The magnetic field detuning spans the range corresponding to $E_\textrm{r}/(k_\textrm{B}T_\textrm{0})$ ranging from 3/2 to 10, which is the interacting, non-unitary regime as defined in Sec.~\ref{Three-body}.
    The dimer break-up rate $\Gamma_\textrm{r}$ (solid line) has a $(B-B_\textrm{0})^{3/2}$ dependence, the inelastic loss rate $\Lambda_\textrm{ad}n_\textrm{0}$ (dotted lines) is constant with $n_\textrm{0}$ the initial density of the three scenarios, the dimer creation rate is $L_\textrm{M} n_\textrm{0}$ (dash-dotted lines), and the loss rate of observer atoms is $\Lambda_\textrm{ad}n_\textrm{D,0}$ (dashed lines) with $n_\textrm{D,0}$ the initial dimer density.}
\end{figure}

We can extract the characteristic rates of the cascade processes appearing in Eqs.~\eqref{eq3} and \eqref{eq4} using the above value of $\Lambda_{\textrm{ad}}$ and Eqs.~\eqref{Gammar} and \eqref{eqnD} of the main text. The values corresponding to the experimental parameters of Fig.~\ref{FigA} are shown in Fig.~\ref{FigB}. The dimer breakup rate $\Gamma_\textrm{r}$ shows a strong magnetic-field dependence while the inelastic atom-dimer collision rate $n_0 \Lambda_{\textrm{ad}}$ is constant. For small magnetic-field detunings, $\Gamma_\textrm{r}$ becomes smaller than $n_0 \Lambda_{\textrm{ad}}$, thus making the consideration of the thermal and non-thermal regimes relevant. In the interacting regime where the loss is well described by the cascade process (see Fig.~\ref{FigA}), we find that $\Gamma_\textrm{r}/(n_0 \Lambda_{\textrm{ad}})$ varies between 0.1 and 7. This indicates that the system can be in a non-thermalized regime or in a transition regime where elastic and inelastic cascade processes happen at similar rates. We also note that in previous studies with  $^6\textrm{Li}$~\cite{PhysRevLett.120.133401, PhysRevA.99.052704},  $\Gamma_\textrm{r}/(n_\textrm{0}\Lambda_{\textrm{ad}})$ was varied from $10^{-1}$ to $10^1$. Furthermore, we note that for the parameters of Fig.~\ref{FigA} and in the approximation of Eq.\,\eqref{eqnD}, the equilibrium dimer density $n_\textrm{D}$ is between 30 and $7\times10^4$ times smaller than the atom density $n$.

With the extracted dimer densities and assuming similar atom-atom and dimer-dimer collisional cross sections, dimer-dimer relaxation processes can indeed be neglected.
Another process to be considered is secondary collisions, which can happen when the fast products of the relaxation collide elastically with trapped atoms or dimers on their way out of the trap.
Following the arguments of the authors of Ref.~\cite{PhysRevA.60.R765}, the loss rate of both dimers and atoms $\gamma_\textrm{sec}$ associated to such collisions can be estimated from 
\begin{equation}
    \tag{A3}
    \gamma_\textrm{sec}\approx 2 \Lambda_\textrm{ad} n n_\textrm{D} b \sigma \exp\boldsymbol{(}b \sigma (n+n_\textrm{D})\boldsymbol{)}
\end{equation}
for a cloud of characteristic size $b$.
Here $\sigma$ is the elastic collisional cross section for a fast collision product (atom or dimer) with an atom or a dimer of the sample.
As the upper limit of the cross section for atom-atom, atom-dimer, and dimer-dimer collisions we choose the inter-atomic elastic cross section at the average collision energy.
This energy is determined by the kinetic energy of the fast product and is on the order of the vibrational deactivation energy.
For the $p$-wave Feshbach resonance of $^6\textrm{Li}$, an estimate for the lower limit of the vibrational deactivation energy is $0.2~\si{\kelvin}$ due to deactivation from the molecular state $X ^1\Sigma^{+}_{g}$ with vibrational quantum number $\nu=38$~\cite{RevModPhys.82.1225} to the molecular state $1 ^3\Sigma^{+}_{g}$ with highest rovibrational level at $4.10~\si{\giga\hertz}$ binding energy~\cite{Abraham}.
Thermal averaging of the elastic cross section then gives an upper bound of $\sigma=5.6\times10^{-18}~\si{\meter\squared}$.
Assuming a cloud size of $b=100~\si{\micro\meter}$, the upper limit for $\gamma_\textrm{sec}$ is approximately $1~\si{\per\second}$ for $B-B_\textrm{0} = 100~\si{\milli\gauss}$ and always several orders of magnitude smaller than $\Lambda_\textrm{ad}n_\textrm{D}$. Thus secondary collisions can be safely neglected.

%--------------------------------------------

\section*{\label{appendixB} Appendix B: Losses in the non-thermal regime}

To study the three-body losses in the non-thermalized regime, we extend Eq.~\eqref{eq4} into a velocity dependent differential equation in the limit of $\Gamma_\textrm{r} \ll n_\textrm{0} \Lambda_{\textrm{ad}}$.
With the approximation of $K_\textrm{M}(E) \propto \delta(E-E_\textrm{r})$ for the dimer creation rate, the phase-space density $\rho(v)$ of the atoms changes with every event where a dimer is created and immediately lost via atom-dimer collision.
The collision energy of two atoms with velocities $\boldsymbol{v}$ and $\boldsymbol{v'}$ is  $E(\boldsymbol{v},\boldsymbol{v'})=m|\boldsymbol{v}-\boldsymbol{v'}|^2/4$, such that the change in phase-space density is:
\begin{equation}
    \tag{B1}
    \label{eq8}
    \begin{aligned}
        \frac{d \rho(|\boldsymbol{v}|)}{dt} =&
        \frac{-\rho(|\boldsymbol{v}|)}{n} \int \,d \boldsymbol{v'} \,d \boldsymbol{v''} K_\textrm{M}\textbf{(}E(\boldsymbol{v'},\boldsymbol{v''})\textbf{)} \rho(|\boldsymbol{v'}|) \rho(|\boldsymbol{v''}|) \\
        &-2\ \int \,d \boldsymbol{v'} K_\textrm{M} \textbf{(}E(\boldsymbol{v},\boldsymbol{v'})\textbf{)} \rho(|\boldsymbol{v}|) \rho(|\boldsymbol{v'}|)
    \end{aligned}
\end{equation}
Here the gas is assumed to be trapped in a box potential and $\rho(v) = n_\textrm{0} \cdot f(E_v)$ is the product of the initial density of atoms with the kinetic energy distribution.
The atom's kinetic energy $E_v = m v^2/2$ is assumed to be independent of the atom's position, and the gas is prepared such that $f(E_v)$ initially is a Maxwell-Boltzmann distribution.

The second term of Eq.~\eqref{eq8} accounts for the process that the atom lost from the distribution with velocity $\boldsymbol{v}$ collides with another atom with velocity $\boldsymbol{v'}$ to form a weakly bound dimer.
In the limit of no thermalization, this dimer is automatically lost from the trap, and thus the loss rate has no additional dependence on the density.
The factor of 2 is due to the interchangeability of the two atoms making up the dimer.
The first term of the equation accounts for the loss of single atoms due to collision with a dimer, formed beforehand by the collision of two atoms of velocities $\boldsymbol{v'}$ and $\boldsymbol{v''}$ .
The velocity group of atoms lost this way depends only on the population of the velocity group and the dimer-creation rate.
Thus no condition is imposed on the velocity of the atom, and the loss scales with the population of the velocity group.

\end{document}